\documentclass[aps,prl,twocolumn,superscriptaddress,altaffilletter,nofootinbib,noeprint]{revtex4-2} 

\usepackage{graphicx} 
\usepackage{soul}
\usepackage{color}
\usepackage[normalem]{ulem}

\usepackage{dcolumn}
\usepackage{bm}

\usepackage{setspace}

\begin{document}

\title{Search for Double Beta Decays of $^{134}$Xe with EXO-200 Phase II}

\title{Search for Double Beta Decays of $^{134}$Xe with EXO-200 Phase II}

\newcommand{\footnoteKurchatov}{Now a division of National Research Center ``Kurchatov Institute,'' Moscow 123182, Russia}
\newcommand{\instStanfordPhys}{Physics Department, Stanford University, Stanford, California 94305, USA}
\newcommand{\instMcGillPhys}{Physics Department, McGill University, Montreal, Quebec H3A 2T8, Canada}
\newcommand{\instKACST}{King Abdulaziz City for Science and Technology, Riyadh, Saudi Arabia}
\newcommand{\instCarletonPhys}{Physics Department, Carleton University, Ottawa, Ontario K1S 5B6, Canada}
\newcommand{\instDukeTUNL}{Department of Physics, Duke University, and Triangle Universities Nuclear Laboratory (TUNL), Durham, North Carolina 27708, USA}
\newcommand{\instIllinoisPhys}{Physics Department, University of Illinois, Urbana-Champaign, Illinois 61801, USA}
\newcommand{\instITEPKurchatov}{Institute for Theoretical and Experimental Physics named by A.I. Alikhanov of National Research Centre ``Kurchatov Institute,'' Moscow 117218, Russia}
\newcommand{\instUSDakotaPhys}{Department of Physics, University of South Dakota, Vermillion, South Dakota 57069, USA}
\newcommand{\instUKentuckyPhys}{Department of Physics and Astronomy, University of Kentucky, Lexington, Kentucky 40506, USA}
\newcommand{\instSLAC}{SLAC National Accelerator Laboratory, Menlo Park, California 94025, USA}
\newcommand{\instTRIUMF}{TRIUMF, Vancouver, British Columbia V6T 2A3, Canada}
\newcommand{\instIHEP}{Institute of High Energy Physics, Beijing 100049, China}
\newcommand{\instWitmem}{Witmem Technology Co., Ltd., No.56 Beisihuan West Road, Beijing, China}
\newcommand{\instLaurentianPhys}{Department of Physics, Laurentian University, Sudbury, Ontario P3E 2C6, Canada}
\newcommand{\instCSUPhys}{Physics Department, Colorado State University, Fort Collins, Colorado 80523, USA}
\newcommand{\instUNCWPhys}{Department of Physics and Physical Oceanography, University of North Carolina at Wilmington, Wilmington, NC 28403, USA}
\newcommand{\instIndianaCEEM}{Physics Department and CEEM, Indiana University, Bloomington, Indiana 47405, USA}
\newcommand{\instDrexelPhys}{Department of Physics, Drexel University, Philadelphia, Pennsylvania 19104, USA}
\newcommand{\instECAPFAU}{Erlangen Centre for Astroparticle Physics (ECAP), Friedrich-Alexander-University Erlangen-N\"urnberg, Erlangen 91058, Germany}
\newcommand{\instTUMUniverse}{Technische Universit\"at M\"unchen, Physikdepartment and Excellence Cluster Universe, Garching 80805, Germany}
\newcommand{\instUVAPhys}{Department of Physics, University of Virginia, Charlottesville, VA 22904, USA}
\newcommand{\instUmass}{Amherst Center for Fundamental Interactions and Physics Department, University of Massachusetts, Amherst, MA 01003, USA}
\newcommand{\instUMDPhys}{Physics Department, University of Maryland, College Park, Maryland 20742, USA}
\newcommand{\instUCBerkeleyPhys}{Department of Physics at the University of California, Berkeley, California 94720, USA}
\newcommand{\instYaleWright}{Wright Laboratory, Department of Physics, Yale University, New Haven, Connecticut 06511, USA}
\newcommand{\instPrincetonPhys}{Department of Physics, Princeton University, Princeton, New Jersey, USA}
\newcommand{\instIBSCUP}{IBS Center for Underground Physics, Daejeon 34126, Korea}
\newcommand{\instSBUPhys}{Department of Physics and Astronomy, Stony Brook University, SUNY, Stony Brook, New York 11794, USA}
\newcommand{\instAlabamaPhys}{Department of Physics and Astronomy, University of Alabama, Tuscaloosa, Alabama 35487, USA}
\newcommand{\instCanonMedical}{Canon Medical Research US Inc., Vernon Hills, IL, USA}
\newcommand{\instIowaStatePhys}{Department of Physics and Astronomy, Iowa State University, Ames, IA 50011, USA}
\newcommand{\instCaltechKellogg}{Kellogg Lab, Caltech, Pasadena, California 91125, USA}
\newcommand{\instBernLHEP}{LHEP, Albert Einstein Center, University of Bern, Bern, Switzerland}
\newcommand{\instDescartesLabs}{Descartes Labs, 100 North Guadalupe, Santa Fe, New Mexico 87501, USA}
\newcommand{\instHamburgIEP}{Institute for Experimental Physics, Hamburg University, 22761 Hamburg, Germany}
\newcommand{\instSNOLAB}{SNOLAB, Sudbury, ON, Canada}
\newcommand{\instLBNL}{Lawrence Berkeley National Laboratory, Berkeley, California, USA}
\newcommand{\instFermilab}{Fermilab, Batavia, IL 60510, USA}
\newcommand{\instSCIPP}{SCIPP, University of California, Santa Cruz, California, USA}
\newcommand{\instUCSDPhys}{Physics Department, University of California, San Diego, La Jolla, California 92093, USA}
\newcommand{\Hawaii}{Department of Physics and Astronomy, University of Hawaii at Manoa, Honolulu, HI 96822, USA}
\newcommand{\windsor}{Department of Physics, University of Windsor, Windsor, ON N9B 3P4, Canada}
\newcommand{\addrLosAngeles}{Los Angeles, California 90025, USA}

\author{S.~Al~Kharusi}
  \altaffiliation{Present address: \instStanfordPhys}
  \affiliation{\instMcGillPhys}

\author{G.~Anton}
  \affiliation{\instECAPFAU}

\author{I.~Badhrees}
  \altaffiliation{Permanent address: \instKACST}
  \affiliation{\instCarletonPhys}

\author{P.S.~Barbeau}
  \affiliation{\instDukeTUNL}


\author{V.~Belov}
  \affiliation{\instITEPKurchatov}
  \altaffiliation{\footnoteKurchatov}

\author{T.~Bhatta}
  \altaffiliation{Present address: \instUKentuckyPhys}
  \affiliation{\instUSDakotaPhys}

\author{M.~Breidenbach}
  \affiliation{\instSLAC}

\author{T.~Brunner}
  \affiliation{\instMcGillPhys}
  \affiliation{\instTRIUMF}

\author{G.F.~Cao}
  \affiliation{\instIHEP}

\author{W.R.~Cen}
  \altaffiliation{Present address: \instWitmem}
  \affiliation{\instIHEP}

% ...

\collaboration{EXO-200 Collaboration}

\date{\today}

\begin{abstract}
EXO-200 was a leading double beta decay experiment consisting of a single-phase, enriched liquid xenon time projection chamber filled with an admixture of 80.672\% \textsuperscript{136}Xe and 19.098\% \textsuperscript{134}Xe. The detector operated at WIPP between 2010 and 2018 and was designed to search for double beta decay of \textsuperscript{136}Xe. Data was acquired in two phases separated by a period of detector upgrades. We report on the search for $0\nu\beta\beta$ and $2\nu\beta\beta$ decay of \textsuperscript{134}Xe with Phase II EXO-200 data, with median 90\% C.L. exclusion sensitivity $T_{1/2}^{0\nu} \geq 3.7\times 10^{23}$ yr and $T_{1/2}^{2\nu} \geq 2.6 \times 10^{21}$ yr, respectively. No statistically significant signal is observed for either decay mode. We set a world-leading lower limit on the half-life of the neutrinoless decay mode of \textsuperscript{134}Xe of $T_{1/2}^{0\nu} \geq 8.7\times10^{23}$ (90\% C.L.) and the second strongest constraint on the two-neutrino decay of $T_{1/2}^{2\nu} \geq 2.9\times10^{21}$ (90\% C.L.), a 3-fold improvement over the EXO-200 Phase I measurement. New constraints are also set for the $2\nu\beta\beta$ and $0\nu\beta\beta$ decays of \textsuperscript{134}Xe to the lowest excited state of \textsuperscript{134}Ba.

\end{abstract}

\maketitle

Double-beta ($\beta\beta$) decay is a rare, second-order weak transition between even-even nuclei with the same mass number and a proton number that differs by two units, which is observable only when single $\beta$ decay is highly suppressed or forbidden by energy and angular momentum conservation. The $2\nu\beta\beta$ decay in which two neutrinos are emitted alongside two electrons is allowed in the Standard Model and has been observed in fourteen isotopes~\cite{ParticleDataGroup:2024cfk}. 
EXO-200 made the first observation of this decay in \textsuperscript{136}Xe~\cite{EXO-200:2011xzf} and precisely measured its half-life as $2.165\pm 0.016\,(stat)\pm 0.059\,(syst) \times 10^{21}$ yr~\cite{EXO-200:2013xfn}. An intriguing possibility is that $\beta\beta$ decay could occur absent the emission of two neutrinos via a process that violates lepton number and would indicate that the neutrino is a massive Majorana fermion, \emph{i.e.}, its own antiparticle~\cite{Schechter:1981bd,Dolinski:2019nrj}. In 2012, EXO-200 set world-leading constraints on the existence of such a decay in \textsuperscript{136}Xe~\cite{EXO-200:2012pdt} and featured competitive sensitivity with its complete data set in 2019~\cite{EXO-200:2019rkq}.

The EXO-200 detector was a single-phase, liquid xenon time projection chamber (LXe TPC) with scintillation light and ionization charge readout. It ran at the Waste Isolated Pilot Plant (WIPP) in New Mexico using an active volume consisting of 175 kg of enriched LXe with a 80.672\% \textsuperscript{136}Xe and 19.098\% \textsuperscript{134}Xe admixture. \textsuperscript{134}Xe is expected to be a $\beta\beta$ emitter, but no $\beta\beta$ decay mode has been observed yet for this isotope. The decay
\begin{equation}
    ^{134}\text{Xe} \rightarrow ^{134}\text{Ba}^{++} + 2 e^{-} + 2 \bar{\nu}_{e}
\end{equation}
to the ground state of \textsuperscript{134}Ba has a $Q$-value of $825.8 \pm 0.9$ keV \cite{Audi:2014eak}.
PandaX-4T holds the most stringent experimental limits on this process to date, reporting half-life lower bounds $T^{2\nu}_{1/2} \geq 2.8 \times 10^{22}$ yr and $T^{0\nu}_{1/2} \geq 3.0 \times 10^{23}$ yr (90\% C.L.)~\cite{PandaX:2023ggs}, which  improve on previous EXO-200 constraints~\cite{EXO-200:2017vqi}.
\textsuperscript{134}Xe can also undergo $\beta\beta$ decay to a 604.7 keV, 2$^+$ excited state of \textsuperscript{134}Ba~\cite{1982BEZY}, which relaxes to the ground state via prompt $\gamma$-ray emission.
A half-life lower bound 
$T^{0\nu}_{1/2}(0^+\rightarrow2^+) \geq 2.6 \times 10^{22}$ yr (90\% C.L.) was set by the DAMA-LXe experiment~\cite{Bernabei:2002ma}.

On phase space grounds alone, $T^{0\nu}_{1/2}$(\textsuperscript{134}Xe) is expected to be $\gg 2.3 \times 10^{26}$ yr, the current lower bound on  $T^{0\nu}_{1/2}(^{136}\mathrm{Xe})$~\cite{KamLAND-Zen:2022tow}. 
The $2\nu\beta\beta$ half-life depends on a phase-space factor $G_{2\nu}$ and a nuclear matrix element (NME) $M_{2\nu}$ with no new physics. $M_{2\nu}$ is, in general, larger for the \textsuperscript{134}Xe transition than for \textsuperscript{136}Xe, $M_{2\nu}(136)/M_{2\nu}(134)\le 1$~\cite{Simkovic:2013qiy,Barea:2015kwa}.
Using $G_{2\nu}(136)/G_{2\nu}(134)=6327$ from  Refs.~\cite{Kotila:2012zza,yale-group} and $T^{2\nu}_{1/2}$(\textsuperscript{136}Xe) from Ref.~\cite{EXO-200:2013xfn}, one can estimate $T^{2\nu}_{1/2}$(\textsuperscript{134}Xe) {$\lesssim  1.4\times10^{25}$} years, possibly within reach of future detectors~\cite{LZ:2021blo}.

The NMEs of the decays to the lowest excited state of \textsuperscript{134}Ba have been calculated in the interacting boson model~\cite{Barea:2015kwa}. With $Q_{\beta\beta}=225$ keV, $G_{2\nu,0\nu}$ are smaller than those for the decays to the ground state, though their values have not been reported. Assuming a $Q^{11}$ scaling for $G_{2\nu}$, $T_{1/2}^{2\nu}(0^+\rightarrow2^+) \sim 10^6\times T_{1/2}^{2\nu}(0^+\rightarrow0^+)$ $\sim 10^{31}$ yr.

EXO-200 detects ionization electrons and scintillation photons produced through energy deposits in the liquid xenon~\cite{Auger:2012gs}. When combined, the anti-correlated nature of these signals greatly improves energy resolution in LXe by minimizing the event-by-event fluctuations of the detected excitation quanta~\cite{EXO-200:2003bso}. EXO-200 exploited this powerful effect to precisely reconstruct events in the few hundred keV to MeV energy range~\cite{EXO-200:2019bbx}. By defining an optimal ``rotated energy" as the best linear combination of charge and light signals, EXO-200 demonstrated an energy resolution $\sigma/E=$ (1.35 ± 0.09)\% (Phase I, 2011-2014) and (1.15 ± 0.02)\% (Phase II, 2016-2018) at 2.615 MeV (\textsuperscript{208}Tl $\gamma$-ray peak)~\cite{EXO-200:2019rkq}.

Events in EXO-200 are characterized by a prompt scintillation signal detected by two arrays of 234 large-area avalanche photodiodes (LAAPDs) ganged into 37 readout channels and placed at both ends of the cylindrical TPC. The scintillation signal determines the event time stamp and provides energy information. For each event, one or more ionization charge ``clusters" are collected on two sets of anodic crossed wire planes placed at opposite ends of the TPC, in front of the LAAPD planes. These provide the 2-D position of each cluster in the plane perpendicular to the detector axis, as well as energy information complementary to that provided by scintillation. 
Two symmetrical electron drift volumes are defined by the anodes and a central cathode electrode. The ``active volume" of the detector is defined as the region bounded by the copper field rings.  
Ionization electrons from each cluster drift with known electric field-dependent velocity for a time measured as the difference between the scintillation time stamp and when charge is collected at the anode, thus providing their depth along the axis of the detector. 

The number of charge clusters in an event defines its multiplicity, which aids discrimination between $\beta$-like and $\gamma$-like events, as $\beta$ decays typically deposit energy in a single cluster (single-site events, SS) while MeV-scale $\gamma$-rays are likely to Compton-scatter, depositing their energy in multiple clusters (multiple-site events, MS).

In EXO-200, the LXe was constantly evaporated and circulated through commercial heated zirconium getters to maintain purity.
Xenon purity was monitored within the TPC by measuring the attenuation of the charge signal \emph{vs.} drift length using $\gamma$-ray calibration sources.
The LXe and TPC were contained within a thin-walled copper vessel immersed in several tonnes of HFE-7000, a hydrofluoroether heat transfer fluid that stays liquid down to LXe temperature. 
The HFE-filled cryostat was cooled with commercial refrigerators and enclosed within a low-radioactivity Pb shield~\cite{Auger:2012gs}. 
The experiment was housed inside a clean room at the WIPP underground facility.
A plastic scintillator muon veto surrounded the detector on four sides of the clean room~\cite{EXO-200:2015edf}.

In Phase II, EXO-200 featured upgraded, lower noise scintillation readout electronics, the LAAPDs were operated at a higher gain than in Phase I ($\sim$200 versus $\sim$100), and the TPC drift field was raised by 50\% to 567 V/cm~\cite{EXO:2017poz}.
Combined, these upgrades yielded improved energy resolution and lower energy threshold, substantially improving sensitivity. Thus, here we present searches for \textsuperscript{134}Xe double beta decays with Phase II data only.

\begin{figure}
    \centering
    \includegraphics[width=0.99\linewidth]{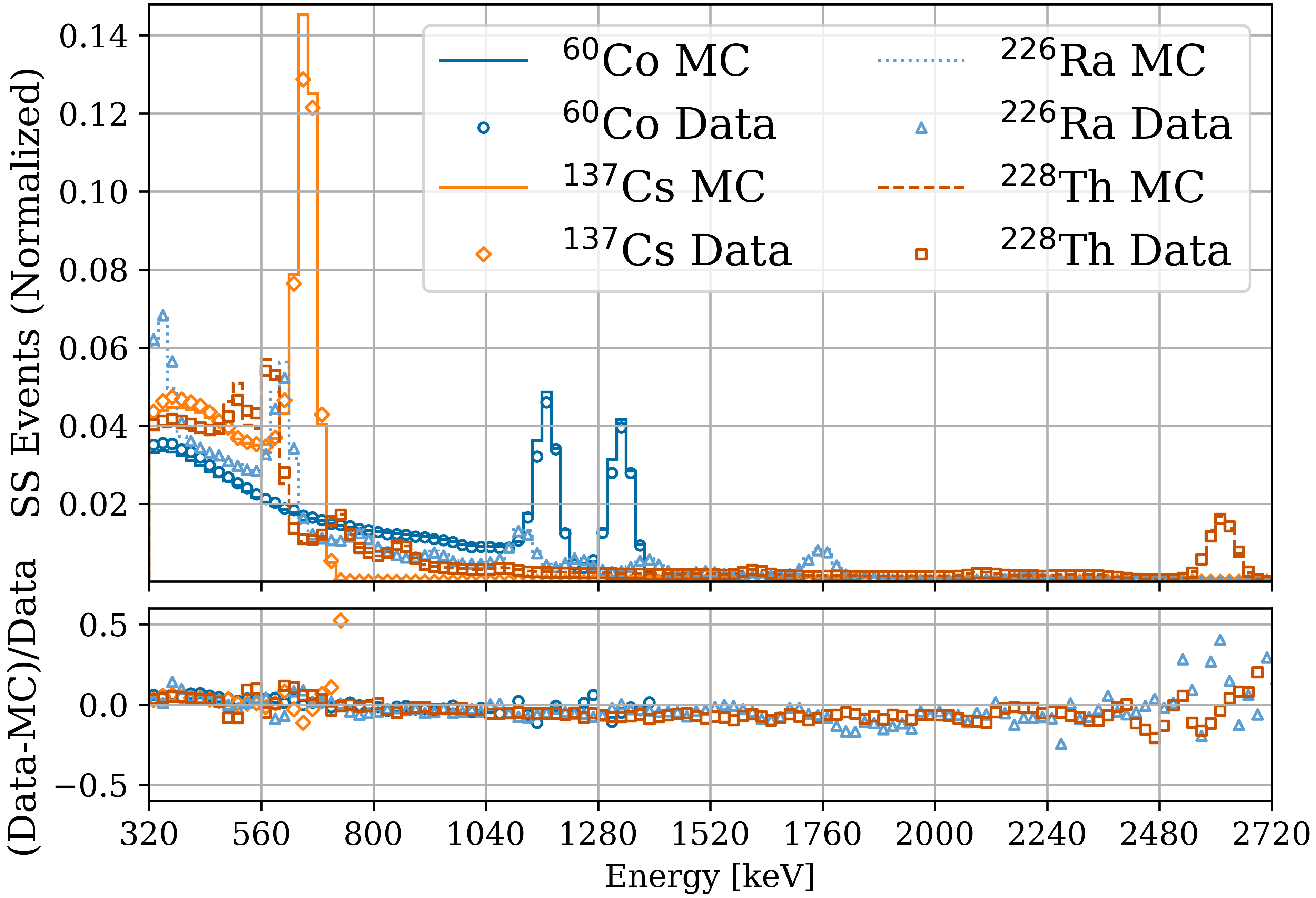}
    \caption{Source agreement in the single-site energy dimension for sources positioned near the cathode. Prominent peaks include 667 keV (\textsuperscript{137}Cs), 1173 and 1332 keV (\textsuperscript{60}Co), 351, 609, and 1764 keV (\textsuperscript{226}Ra), and 2615 keV (\textsuperscript{228}Th).}
    \label{fig:SourceAgreement}
\end{figure}

The detector is calibrated by deploying known gamma emitters (\textsuperscript{137}Cs, \textsuperscript{60}Co, \textsuperscript{226}Ra, and \textsuperscript{228}Th) at various positions along a guide tube surrounding the detector~\cite{EXO-200:2013xfn} and comparing detector response against simulations of these sources. We include 
data taken with a \textsuperscript{137}Cs source as a calibration peak below $Q_{\beta\beta}$ of $^{134}$Xe.  The agreement between the measured single-site energy spectra of the calibration sources and simulations is shown in Fig.~\ref{fig:SourceAgreement}. 
The LXe purity is monitored through measurements of the \textsuperscript{208}Tl 2.6 MeV $\gamma$ peak from the decay chain of \textsuperscript{228}Th.

The analysis procedure is similar to previous EXO-200 searches. 
After basic quality cuts, selection cuts are imposed on the raw data to minimize background while preserving signal efficiency. Events are selected within a fiducial LXe volume whose boundaries are sufficiently distant from all detector materials to eliminate surface $\alpha$ and $\beta$ backgrounds. Events are classified by multiplicity and energy. A likelihood fit to the data that simultaneously includes the energy, multiplicity, and standoff distance (\emph{i.e.}, the spatial separation between a reconstructed charge cluster and the nearest detector surface) is performed against a background model composed of a mixture of simulated components to test for the presence of excess events compatible with the signals.

The selection cuts in the searches for decays to the $^{134}$Ba ground state are identical to those described in Ref.~\cite{EXO-200:2019rkq} except that only events for which the position of all charge clusters is fully reconstructed are selected, which more tightly constrains multi-site backgrounds.
To increase the detection efficiency of the $0^+\rightarrow2^+$ decay signal, a mostly multi-site event, we require only a partial 60\% 3-D reconstruction as in Refs.~\cite{EXO-200:2015edf,EXO-200:2019rkq} with a dedicated calibration built for these events. A muon veto cut is applied as described in Refs.~\cite{EXO-200:2013xfn,EXO-200:2015edf}. After muon and coincidence cuts, the remaining livetime for this analysis is 587.6 days, corresponding to a total fiducial volume \textsuperscript{134}Xe exposure of 28.5 kg-yr (212.8 mol-yr). 

We opt not to use machine-learning classifiers that were developed for previous EXO-200 searches~\cite{EXO:2018bpx,EXO-200:2019rkq}. These tools enhance separation between single and multi-site events, which is used to discriminate against $\gamma$-like backgrounds. Since the primary backgrounds to the searches presented here are topologically similar to the signals, the use of a machine learning discriminator is not justified. 

The fit to data is performed using a maximum likelihood estimator with probability distribution functions (PDFs) built from simulated signal and backgrounds. The normalization of the best fit model is allowed to float as a systematic uncertainty, which is additive to the negative log-likelihood. PDFs are generated using EXO-200 software based on GEANT4~\cite{Allison:2006ve}. The negative log likelihood (NLL) function, modeling of the detector geometry, and simulated signal generation are described in Ref.~\cite{EXO-200:2013xfn}.

Rotated energy, standoff distance, and SS fraction are fit simultaneously. PDFs are constructed by smearing simulated spectra with a resolution model based on calibration data. In the absence of statistically significant signal, half-life lower limits are calculated by profiling the NLL as a function of signal counts. For 90\% confidence level (C.L.), $\Delta NLL$ = 1.35. 

When the measured signal count is close to a physical boundary, this can impart a bias on the statistical quantities by underestimating the confidence interval, as well as from incorrectly accounting for statistical fluctuations in bins near the region of interest.
To account for biases from being near a physical limit, we adopt the method of Feldman and Cousins~\cite{Feldman}, often used in similar searches ~\cite{Majorana:2022udl}.
Maximum likelihood fits are performed on multiple instances of Monte Carlo resampling of the measured background with injection of known amounts of signal. The range of signal allowed by the dataset is that defined by including 90\% of results from the toy MC simulations. We find that this procedure recovers Wilks' theorem and our log likelihood profiles become $\chi^{2}$ distributed.

\begin{table}[h]
    \centering
    \resizebox{\columnwidth}{!}{
    \begin{tabular}{|c|c|c|}
    \hline
        \bf Constraint & \bf $2\nu\beta\beta$ $(2^+)$ & \bf $0\nu\beta\beta$ $(2^+)$ \\
        \hline
        Single-Site Fraction & 3.4\% (4.4\%) & 3.4\% (4.4\%)\\
        \hline
        Event Rate Norm. & 3.4\% (3.3\%) & 3.4\% (3.3\%)\\
        \hline
        Signal-Specific  & $a=18.3\%$ $(44.6\%)$ & $a=16.3\%$ $(28.2\%)$ \\
        Normalization & $b=4411$ $(514)$ cts & $b=16$ ($57$) cts \\
        \hline
        Neutron Capture & 10\% & 10\% \\
        Fraction~\cite{EXO-200:2015edf} & & \\
        \hline 
        Radon in the LXe~\cite{EXO-200:2013xfn} & 20\% & 20\% \\
        \hline
    \end{tabular}
    }
    \caption{Summary of constraints added to the fit with reference to their source. Details of the calculation of single-site fraction, event rate normalization and signal-specific normalization fraction are discussed in the text. Values in parentheses refer to decays to the  $^{134}$Ba excited state.}
    \label{tab:constraints}
\end{table}

The same five systematic uncertainties that were included in the Phase I search~\cite{EXO-200:2017vqi}, listed in Table~\ref{tab:constraints}, are either calculated or drawn from previous analyses and added as Gaussian constraints on the NLL fit. 
The discrepancy in the fraction of single-site events ($\frac{SS}{SS+MS}$) is calculated as described in Ref.~\cite{EXO-200:2017vqi}. (This uncertainty is evaluated separately for the searches for decays to the ground and excited states due to their different 3-D reconstruction requirement).

\begin{figure*}[t!]
    \centering
    \includegraphics[width=0.8\linewidth]{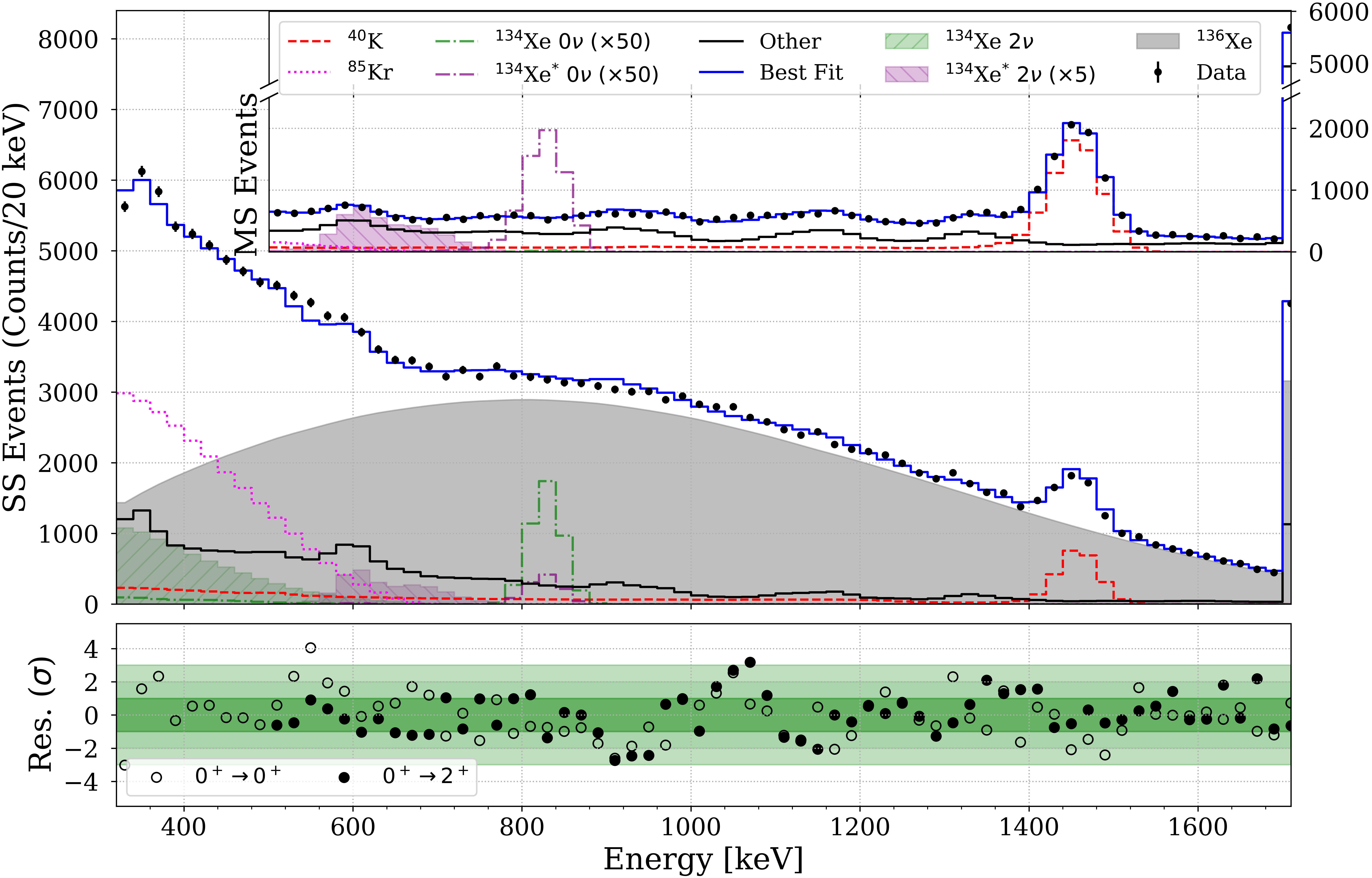}
    \caption{Best fit to the single-site (SS, main panel) and multi-site (MS, inset) energy spectra, shown over the energy range used for the analysis. The MS dataset includes events with partial 3D spatial reconstruction, see text for details. The fits to both $2\nu$ and $0\nu$ $\beta\beta$ decays to the ground state (GS) and excited state (ES) of $^{134}$Ba all return zero signal counts with $\chi^{2}_{red,SS}=1.57$ and $\chi^{2}_{red,MS}=1.09$, respectively. The expected spectra of the $\beta\beta$ signals are shown as follows: i) at the measured half-life lower limit ($2\nu$ GS mode, shaded), ii) magnified 5-fold (2$\nu$ ES mode) and iii) magnified 50-fold to make them visible ($0\nu$ modes, line-dotted).}
    \label{fig:fit}
\end{figure*}

Uncertainties in the background model (detailed in Ref.~\cite{EXO-200:2013xfn} and summarized in the caption of Fig.~\ref{fig:fit}) may affect the signal rate. We quantify the effect on the fit by changing the location of certain background components. The effect of subdominant components excluded from the background model is also treated as a systematic uncertainty. These are \textsuperscript{60}Co in the source tube and inner cryostat materials, \textsuperscript{214}Bi in the thin air gap between the cryostat and the lead shield, \textsuperscript{232}Th and \textsuperscript{238}U in the HFE and the cryostat, as well as the unlikely presence of \textsuperscript{88}Y in the LXe. \textsuperscript{39}Ar and \textsuperscript{210}Bi in the LXe were estimated to be negligible and excluded.   

The signal-specific normalization measures small differences between the data and simulation, and is quantified with the same method as in Ref.~\cite{EXO-200:2017vqi}. We use a similar reweighting scheme to previous analyses. However, 
to more accurately model the low energy range, we include \textsuperscript{137}Cs source data which has not previously been used. Other low-energy backgrounds are re-weighted using \textsuperscript{60}Co.

The uncertainties in the signal shape are folded into the fit as a conditional constraint on the normalization of the signal PDF. The normalization is allowed to float around unity in a range defined by $\sigma=\sqrt{(a\cdot N)^2+b^2}$, where $a$ is the percent error from the background model uncertainty, $b$ is the bias defined above, and $N$ is the number of signal counts.

The constraints listed in Table~\ref{tab:constraints} represent systematic analysis choices and are added to the likelihood fit as multiplicative gaussians which widen the likelihood profile. The impact of each constraint is quantified as a change on the upper limit and is recorded in Table~\ref{tab:systematics}.

\begin{table}[h]
    \centering
    \begin{tabular}{|c|c|c|}
    \hline
        \bf Constraint & \bf $2\nu\beta\beta$ ($2^+$) & \bf $0\nu\beta\beta$ ($2^+$) \\
        \hline
        SS Fraction & $<1\%$ ($<1\%$) & 15.4\% (29.8\%)\\
        \hline
        Event Rate Norm. & $<1\%$ ($<1\%$) & 11.3\% ( $<1\%$ )\\
        \hline
        Signal-Specific Norm. & 85.3\% (90.2\%) & 4.1\%  (32.4\%)\\
        \hline
        Neutron Capture Frac. & $<1\%$ ($<1\%$) & 1\% ( $<1\%$ )\\
        \hline 
        Radon in the LXe & $<1\%$ ($<1\%$) & 13.4\% ( $<1\%$ )\\
        \hline
    \end{tabular}
    \caption{Summary of the contributions of each fit constraint to the reported upper limit on the number of counts of each decay mode. Values in parentheses are in reference to the decays to the excited state.}
    \label{tab:systematics}
\end{table}
 
As noted in Ref.~\cite{EXO-200:2017vqi}, the spectra of the $\beta$ decay of \textsuperscript{85}Kr and the $0^+\rightarrow0^+$ $2\nu\beta\beta$ decay of \textsuperscript{134}Xe are highly degenerate. 
Fits with and without constraints were scanned over \textsuperscript{85}Kr and \textsuperscript{134}Xe counts. After the systematic uncertainties (dominantly the signal normalization error in this case) are incorporated, we see no anti-correlation between signal and background counts~\cite{supplemental}.  

To evaluate the median sensitivity for each of the signals, we perform a fit to the data assuming a (signal-specific) background-only hypothesis, \emph{i.e.}, enforcing no signal in the fit. 
The resulting background-only fit is then resampled through bootstrapping to generate a large group of datasets, each undergoing a fit where the signal is allowed to float as a measure of the impact of fluctuations in the data on the best-fit of the signal. The median 90\%~C.L. upper limit of all these fits is taken as the median sensitivity. 

The 90\%~C.L. median sensitivities for the $0^+\rightarrow0^+$ decays are {$T_{1/2}^{0\nu}=3.7\times 10^{23}$ yr} and {$T_{1/2}^{2\nu}=2.6\times 10^{21} $ yr}. The 2-fold improvement over the sensitivity for the $0\nu$ decay mode in Phase I~\cite{EXO-200:2017vqi} is due to better modeled low energy backgrounds and improved energy resolution. A seven-fold increase in sensitivity is obtained for the $2\nu$ decay mode thanks to a higher signal acceptance allowed by a substantially decreased energy threshold.

The fit to both $0^+\rightarrow0^+$ $\beta\beta$ decay modes returns a best value of zero signal counts after all uncertainties are accounted for and independently of the chosen energy threshold.
The fit to the dataset is shown in Fig.~\ref{fig:fit}, using the selected threshold of 320 keV. Absent statistically significant signals, lower bounds on the half-life for each decay mode are computed as
\begin{equation}
    T_{1/2} \geq \frac{ln(2)\cdot t_{exp}\cdot \epsilon}{N_{U.L.}} \cdot \frac{N_A}{0.1339},
\label{eq:half-life}
\end{equation}
where $t_{exp}$ 
is the exposure 
in kg$\cdot$yr, $\epsilon$ the event detection efficiency, $N_A$ Avogadro's number, and $N_{U.L}$ the 90\%~C.L. upper limit on the number of signal events. 0.1339 is the molar mass of \textsuperscript{134}Xe.

The Feldman-Cousins procedure discussed above yields a 90\%~C.L. upper limit of 102 \textsuperscript{134}Xe $0\nu\beta\beta$ decay events. With a 99.1\% event detection efficiency, this translates to a lower limit on the half-life of this decay of
\begin{equation}
    T_{1/2}^{0\nu}(0^+\rightarrow0^+) \geq 8.7 \times 10^{23} \, \text{yr} \,\,\,\,\,\,\,\,\,\,\,\,\,\, 90\% \,\text{C.L.}
\end{equation}
a 2.9-fold improvement over existing constraints~\cite{PandaX:2023ggs} and a 7.9-fold improvement over the EXO-200 Phase I result~\cite{EXO-200:2017vqi}.

With an event detection efficiency of 25.5\% and 90\%~C.L. upper limit of 7606 counts, the lower limit on the \textsuperscript{134}Xe $2\nu\beta\beta$ decay half-life is
\begin{equation}
    T_{1/2}^{2\nu}(0^+\rightarrow0^+) \geq 2.9 \times 10^{21} \, \text{yr} \,\,\,\,\,\,\,\,\,\,\,\,\,\, 90\% \,\text{C.L.}
\end{equation}
a three-fold improvement over the EXO-200 Phase I result~\cite{EXO-200:2017vqi}, but weaker than existing constraints~\cite{PandaX:2023ggs}.

The \textsuperscript{136}Xe $2\nu\beta\beta$ decay background is used to check the reliability of the background model by leaving its spectral component unconstrained in the fit and comparing the best-fit result with its known half-life. We measure $T_{1/2}^{2\nu\beta\beta}(^{136}\text{Xe}) = 2.23\pm0.1\,\mathrm{(stat.)}\times10^{21}$ yr based on the fit shown in Fig.~\ref{fig:fit}, in excellent agreement with the previous EXO-200 result~\cite{EXO-200:2013xfn}.

The fit to both $0^+\rightarrow2^+$ modes was carried out using a low energy threshold of 500 keV and returned no signal counts. The sensitivity for this analysis does not improve below this threshold. To account for the possibility of detecting events originating outside the fiducial volume, the event detection efficiency is multiplied by a correction factor determined from simulations of the active volume. The detection efficiency of $2\nu\beta\beta$ ($0\nu\beta\beta$) decays to the \textsuperscript{134}Ba excited state is 70.4\% (68\%). We calculate a 90\% C.L. median sensitivity of $T_{1/2}^{2\nu}\ge4.7\times 10^{22}$ yr and $T_{1/2}^{0\nu}\ge 3.5\times 10^{23}$ yr. The 90\% C.L. upper limit on the number of $2\nu\beta\beta$ ($0\nu\beta\beta$) counts is 1243 (151), respectively, yielding the following lower bounds on their half-lives:
\begin{equation}
    T_{1/2}^{2\nu}(0^+\rightarrow2^+)\ge5.1\times 10^{22}  \, \text{yr} \,\,\,\,\,\,\,\,\,\,\,\,\,\, 90\% \,\text{C.L.}
\end{equation}

the first reported measurement of this decay, and
\begin{equation}
    T_{1/2}^{0\nu}(0^+\rightarrow2^+)\ge 4.0\times 10^{23}  \, \text{yr} \,\,\,\,\,\,\,\,\,\,\,\,\,\, 90\% \,\text{C.L.}
\end{equation}
a 
fifteen-fold improvement over existing constraints~\cite{Bernabei:2002ma}.

This work was supported by grants from the NSF (PHY-2111213 and PHY-2514767) as well as from the DOE (DE-SC0020509). EXO-200 was supported by the DOE (DE-SC0017970, DE-FG02-01ER41166, DE-SC0021383, DE-SC0014517) and NSF in the U.S., the NSERC in Canada (CRC-2019-00399, SAPPJ-2017-00029), the SNF in Switzerland, the IBS in Korea, the RFBR (18-02-00550) in Russia, the DFG in Germany, and CAS and ISTCP in China. The EXO-200 data analysis and simulation uses resources of the SLAC Shared Scientific Data Facility (S3DF). We gratefully acknowledge the KARMEN collaboration for supplying the cosmic-ray veto detectors, as well as the WIPP for their hospitality.

\bibliography{references.bib}

\end{document}